\documentclass[useAMS,usenatbib]{mn2e}
\usepackage{times}
\usepackage{amsmath}
\usepackage{graphicx}
\usepackage{aas_macros}

\begin{document}

\title[Generalized three body problem]
{Generalized three body problem and the instability of the core-halo objects in binary systems}

\author[Andrzej Odrzywolek]{Andrzej Odrzywolek\thanks{E-mail:
andrzej.odrzywolek@uj.edu.pl}\\
M. Smoluchowski Institute of Physics, 
 Jagiellonian University,  
 Reymonta 4, 
 30-059 Krakow, 
 Poland
}

\date{\today}

\maketitle

\begin{abstract}
Goal of the presented research is to construct simplified model of the core-halo structures
in binary systems. Examples are provided by Thorne-Zytkov objects, hot Jupiters, protoplanets with large moons, 
red supergiants in binaries and globular clusters with central black hole. 
Instability criteria due to resonance between internal and orbital frequencies in such a systems
has been derived. To achieve assumed goals, generalized planar circular restricted three body problem 
is investigated with one of the point masses, $M$, replaced with spherical body of finite size.
Mechanical system under consideration includes two large masses $m$ and $M$ 
and the test body with small mass $\mu$. Mass $\mu$, initially, is placed in the 
geometric center of mass $M$, and shares its orbital motion. 
Only gravitational interactions are considered, and non-point mass $M$ is assumed to be rigid with 
rotational degrees of freedom neglected. Equations of motion are presented, 
and linear instability criteria are derived using quantifier elimination.

Motion of the test mass $\mu$ is shown to be unstable due to resonance between orbital and internal frequencies
if $\frac{M}{d^3} < \frac{4}{3} \pi \rho < \frac{ M + 3 m \left( 1+\mu/M \right)^{-1}}{d^3}$,
where $\rho$ is the central density of mass $M$, and $d$ distance 
between masses $m$ and $M$ (circular orbit diameter). In the framework of model, 
the central mass $\mu$ can be ejected if 
resonance conditions are met during the evolution of the system.
% conclusions
The above result is important for core-collapse supernova theory, 
with mass $\mu$ identified with helium core of the exploding massive star. The instability 
cause off-center supernova ''ignition''  relative to the center-of-mass of the hydrogen envelope.
The instability is also inevitable during protoplanet growth, 
with hypothetical ejection of the rocky core from gas giants and
formation of the ''puffy planets'' due to resonance with orbital frequency. 
Hypothetical central intermediate black holes of the globular clusters are also 
in unstable position with respect to perturbations caused by the Galaxy. For sake of curiosity I note, 
that the Earth-Moon, and the Earth-Sun systems are stable in the above sense, 
with test body $\mu$ being the artificial black hole created in the 
failed high-energy physics experiment.
\end{abstract}


\section{Introduction}

Traditional classical mechanics approach to astrophysical binaries usually assume 
that mechanical system can be described in terms of point masses. 
However, in many important situations, this assumption is broken, e.g., 
the motion of the spacecraft in the gravitational field
of the non-spherical asteroid \citep{wang2013gravity}, 
require more general description \citep{2012A&A...541A.130H}. In astrophysics, we frequently have to deal with 
core-halo objects. These bodies still possess spherical symmetry. 
Total mass is unevenly divided into nearly point-like
central object, and extensive, low-density envelope.

For example, all red-giants are composed of small, high-density helium core, surrounded by
a huge low-density envelope \citep{2002RvMP...74.1015W}. Noteworthy, mass of the envelope is usually dominant,
and vary by order of magnitude from few solar masses up to $\sim$100~M$_\odot$. 
On the contrary, mass of the helium core is roughly 4-5~M$_\odot$ for all stars.

Gaseous giant planets (''Jupiters'') are composed of the rocky core and extended envelope.
Inside icy moons (e.g. Europa, Enceladus) and exoplanets 
(''blue ocean'' super-earths, \citealp[see][]{2011ConPh..52..403H}) 
we also find rocky core, this time ''floating'' in the surrounding
liquid ocean  \citep{2003ApJ...596L.105K,2004Icar..169..499L}. Therefore, in some situations, 
aforementioned bodies should be treated as two-component
structures.

Globular cluster (GC) with 
central intermediate-mass black hole 
\citep[IMBH, see][]{2013ApJ...768...26U, 2013ApJ...776..118S, 2013A&A...554A..63F, 
2013A&A...558A.117L,2013MmSAI..84..129L,2013A&A...552A..49L, 2013A&A...555A..26L} is another
very important case of the core-halo object. 
Noteworthy, non-gravitational effects are essentially negligible in GC-IMBH
system, providing perfect testbed for presented theory.

There are also more exotic examples. \citet{1977ApJ...212..832T} object, composed
of central neutron star or black hole and normal star provides 
illustrative case \citep{1999ApJ...514..932V,1993MNRAS.263..817C,1992ApJ...386..206C}. 
Yet another example is the non-standard Solar model with black hole inside 
\citep{1971MNRAS.152...75H,1975ApJ...201..489C}. 
Artificial black hole created on Earth in failed high-energy experiment 
\citep{2010JHEP...02..079C,2010JPhCS.237a2008B,2010PhRvD..81e7702G}
also would lead to creation of the core-halo system.

The big question is, what if the core-halo system is a part of orbiting gravitationally bound system? 
For red giants, it would be a binary companion star. For planets: moons,
and \textit{vice versa}. For globular clusters, it is the host Galaxy. In all above examples,
binary interaction effects are known to be non-negligible, e.g., 
tidal interactions. In extreme cases, mass transfer or a total
disruption of one component is possible. Are there other types of instability, resonant in particular?
I try to answer this question in the framework of mechanical model.

Original motivation for creation of presented model is suggestion of \cite{2011ApJ...733...78A}, 
that in core-collapse supernovae iron core might be displaced with respect to geometrical symmetry 
center of the extended and usually much more massive hydrogen envelope.
\citet{2011ApJ...733...78A} proposed hydrodynamical L=1 instabilities during Si burning as a main cause 
of the displacement. 
 \cite{2011ApJ...733...78A} wrote: \textit{
,,If there were a driving mechanism for core-mantle oscillation, here would be an asymmetry due 
to the displacement of the core and mantle relative to the center of mass.''}
Here I propose another mechanism: gravitational instability in binary core-halo system.

The point mass in the center of an extended object would oscillate, if it was perturbed
(displaced) from central position. Is this position stable with respect to perturbations caused 
by the third body orbiting outside? 
Naively, one might expect, that it is always possible to tune internal frequency 
and orbital frequency. However, we are dealing
with energy-conserving system. Many years of struggle to answer very similar
questions in the classic three body problem \citep{2008LNP...760...59M} 
suggest caution, and rigorous mathematical approach.

The article is organized as follows. In Sect.~\ref{model}
details of the  model created to describe core-halo
system in binary are presented. Approach based on restricted planar circular three body problem
(thereafter RPCTBP) has been presented in Subsect.~\ref{rpc3bp_sect},
together with linear stability analysis for uniform density ball.
Subsection~\ref{Sect_numver} contains numerical verification
of the results, and behavior in non-linear regime. 
In Subsect.~\ref{instability_3d} much more general model with non-planar motion
and third body of finite mass $\mu$ is provided. Potential
astrophysical sites of the instability are presented in Section~\ref{sites}:
massive pre-supernova stars (Subsect.~\ref{Subsect:redsupergiants}), 
(exo)planets and moons (Subsect.~\ref{Subsect:exomoons})
and globular clusters with IMBH (Subsect.~\ref{Subsect:GC}). Importance of instability, 
limitations of mechanical model, chances
for observational verification and directions of future research are
summarized in concluding Sect.~\ref{final}.

\section{Derivation and numerical verification of the instability criteria}
\label{model}

\subsection{Restricted planar circular three body problem approach \label{rpc3bp_sect}}

\begin{figure*}
\includegraphics[width=\textwidth]{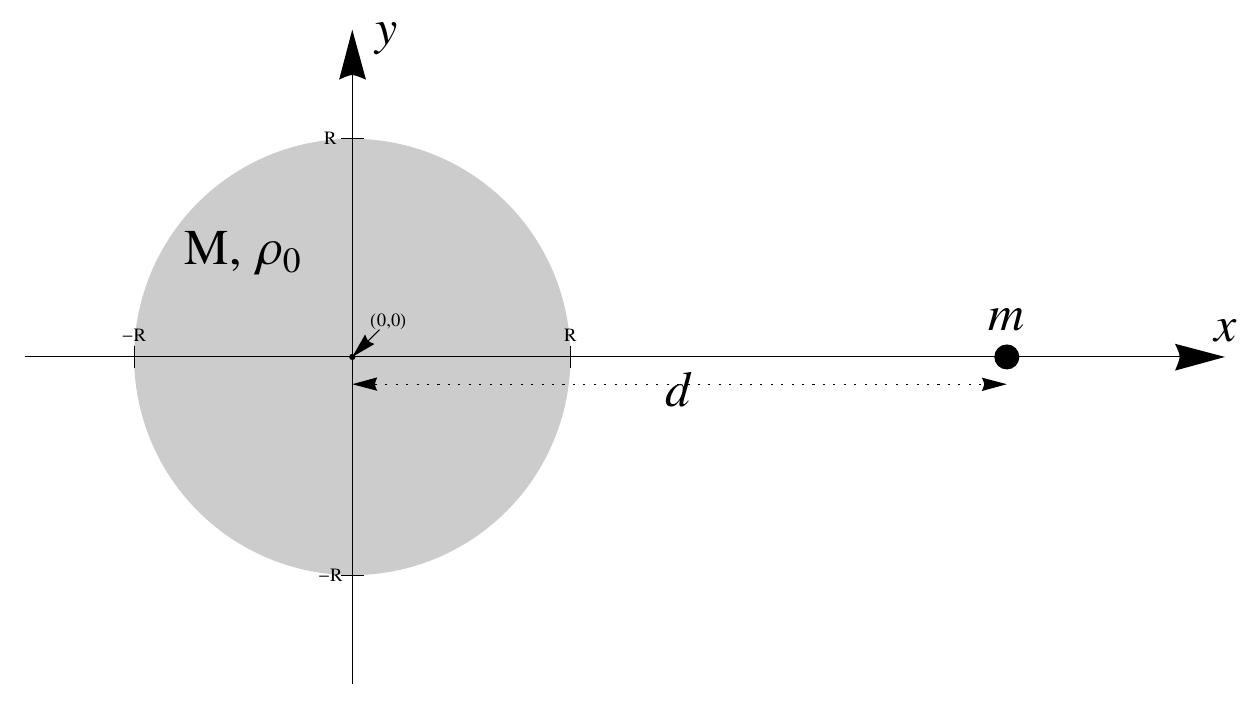}
\caption{\label{coordSYS} Co-rotating coordinate system used to derive \eqref{rpc3bsys}.}
\end{figure*}

To handle dynamics of the core-halo object in binary system, the following model has been created.
Mechanical system of interest (cf.~Fig.~\ref{coordSYS}) includes two masses $m$ and $M$ orbiting the center of mass on circular orbits.
Mass $m$ (first body) is a point mass. Mass $M$ (second body) is extended spherical body 
with known density $\rho(r)$. The body with mass $M$ is rigid.
Rotational degrees of freedom for mass $M$ are not considered, though. Third body is a test body, 
so its mass is assumed to be negligible (see Subsect.~\ref{instability_3d} for more general model
with this assumption relaxed).
In this subsection only, in order to simplify formulae, and facilitate
derivation of linear instability criteria, I further assume that density $\rho(r)$ inside mass $M$ is 
constant and equal $\rho$. 
Additionally, mass $m$ is assumed to orbit outside radius $R$ of mass $M$. This allows
us to use classic RPCTBP formulae.
Distance between geometrical center of mass $M$ and $m$ is equal to $d$, and $d \geq R$. 
Third test body is initially in the center of mass $M$, where also coordinate origin is placed. 
In co-rotating cartesian system $(x,y)$ (cf. Fig.~\ref{coordSYS}), 
equations of motion for third body, restricted to orbital plane, are:
\begin{subequations}
\label{rpc3bsys}
\begin{align}
\ddot{x} -2\, \omega\, \dot{y} - k \, x + & \frac{ G\, m\, (x-d)}{\left[ (x-d)^2+y^2 \right]^{3/2}}
+ \frac{G\, m}{d^2}=0,\\
%\end{equation}
%\begin{equation}
\ddot{y} +2\, \omega\, \dot{x} + k \, y + & \frac{ G\, m\, y}{\left[ (x-d)^2+y^2 \right]^{3/2}	}
=0,
\end{align}
\end{subequations}
where dot denotes time derivative, and, from Kepler law: 
\begin{equation}
\label{kepler_frequency}
\omega^2 = \frac{G\, (m+M)}{d^3}
\end{equation}
and:
\begin{equation}
\label{k}
k =\frac{4}{3} \pi G \rho -\omega^2 .
\end{equation}

System \eqref{rpc3bsys} is very similar to the classical planar restricted circular 
three body problem, see \cite{0951-7715-24-5-002}. System \eqref{rpc3bsys}, without 
terms \textit{explicite} involving gravitational constant $G$,  describe Foucault pendulum problem \citep{1969mech.book.....L}.

Conserved energy for system \eqref{rpc3bsys} is:
\begin{subequations}
\begin{equation}
E = \frac{1}{2} \dot{x}^2 + \frac{1}{2} \dot{y}^2 + U(x,y),
\end{equation}
\begin{equation}
\label{energy}
U = 
\frac{1}{2} k \, 
\left(  x^2+y^2\right) - \frac{ G\, m }{\sqrt{(x-d)^2+y^2}}+\frac{G\, m (x+d)}{d^2}
\end{equation}
\end{subequations}
Equations of motion \eqref{rpc3bsys} and energy \eqref{energy} with constant $k$ are suitable for 
analysis of the small (linear) perturbations of the test body only.
To explore non-linear effects, possible ejection of the test body from mass $M$ in particular,  
we have to use model with non-constant density $\rho(r)$. This is done in Sect.~\ref{Sect_numver}.

Linearization of the equations \eqref{rpc3bsys} for small perturbations around point $x=0, y=0$
is done as follows.  After substitution $x(t) = \epsilon\, \zeta(t), 
y(t) = \epsilon\, \xi(t)$ into \eqref{rpc3bsys}, series expansion has been calculated with respect to $\epsilon$, 
and higher-order terms dropped. Following linear system has been obtained:
\begin{subequations}
\label{linearized}
\begin{equation}
\ddot{\zeta}-2\, \omega\, \dot{\xi}+ (k-2q)\, \zeta =0,
\end{equation}
\begin{equation}
\ddot{\xi} + 2\, \omega\, \dot{\zeta} + (k+q)\, \xi =0,
\end{equation}
\end{subequations}
where:
\begin{equation}
\label{q}
q = \frac{G\, m}{d^3}.
\end{equation}
Eigenvalues $\lambda$ of the system \eqref{linearized} are solutions to the algebraic equation:
\begin{equation}
\label{eigenvalues}
( \lambda^2+k-2q)(\lambda^2 + k + q) + 4\, \omega^2 =0.
\end{equation}

System is considered linearly unstable with respect to small perturbations if at least 
one solution of \eqref{eigenvalues} has a positive real part:
$$
Re(\lambda)>0.
$$
Resolving above conditions using quantifier elimination \citep{strzebonski, Liska01011993} 
lead us to the instability criteria:
\begin{subequations}
\label{crit}
\begin{equation}
\label{crit1}
\frac{M}{d^3} < \frac{4}{3} \pi  \rho < \frac{M + 3\, m}{d^3},
\end{equation}
\begin{equation}
\label{crit2}
\frac{4}{3} \pi  \rho < \frac{1}{2} \frac{m}{d^3} \frac{M-m/8}{M+m}.
\end{equation}
\end{subequations}

Criteria \eqref{crit} has been verified solving \eqref{rpc3bsys} numerically, 
and solving linearized system \eqref{linearized} analytically. Result
\eqref{crit1} can be obtained from analysis of the potential \eqref{energy} extremum
at $x=0,y=0$ as well.  Left-hand side of \eqref{crit1}, i.e., condition $ M/d^3 <4/3 \pi \rho$, 
is trivial from astrophysical point of view, because the central density cannot 
be smaller than average density. In any realistic astronomical body, 
density decrease outwards: $\rho(r) \leq \rho(0)$. The same argument 
apply to \eqref{crit2}. Even in the most favorable
situation $m=2M$, the central density should be less than half of the average density 
for mass $m$ for this instability to occur\footnote{However, artificial body
with such properties could be created, and experiments performed
in micro-gravity at orbital station.}.

If we assume, that density decreases outwards
from the center,  we may simply write simplified form of \eqref{crit}, 
relevant to the astrophysical applications:
\begin{equation}
\label{crit_astro}
\frac{4}{3} \pi  \rho < \frac{M + 3\, m}{d^3}.
\end{equation}

It is illustrative, to compare instability \eqref{crit_astro}
with Roche limit \citep{1963ApJ...138.1182C, 2003ApJ...588..509G}.
Let rewrite \eqref{crit_astro} as:
\begin{equation}
\label{roche_like}
\frac{d}{R} < \left( 1+ \frac{3m}{M} \right)^{1/3}.
\end{equation}

\begin{figure}
\includegraphics[width=0.5\textwidth]{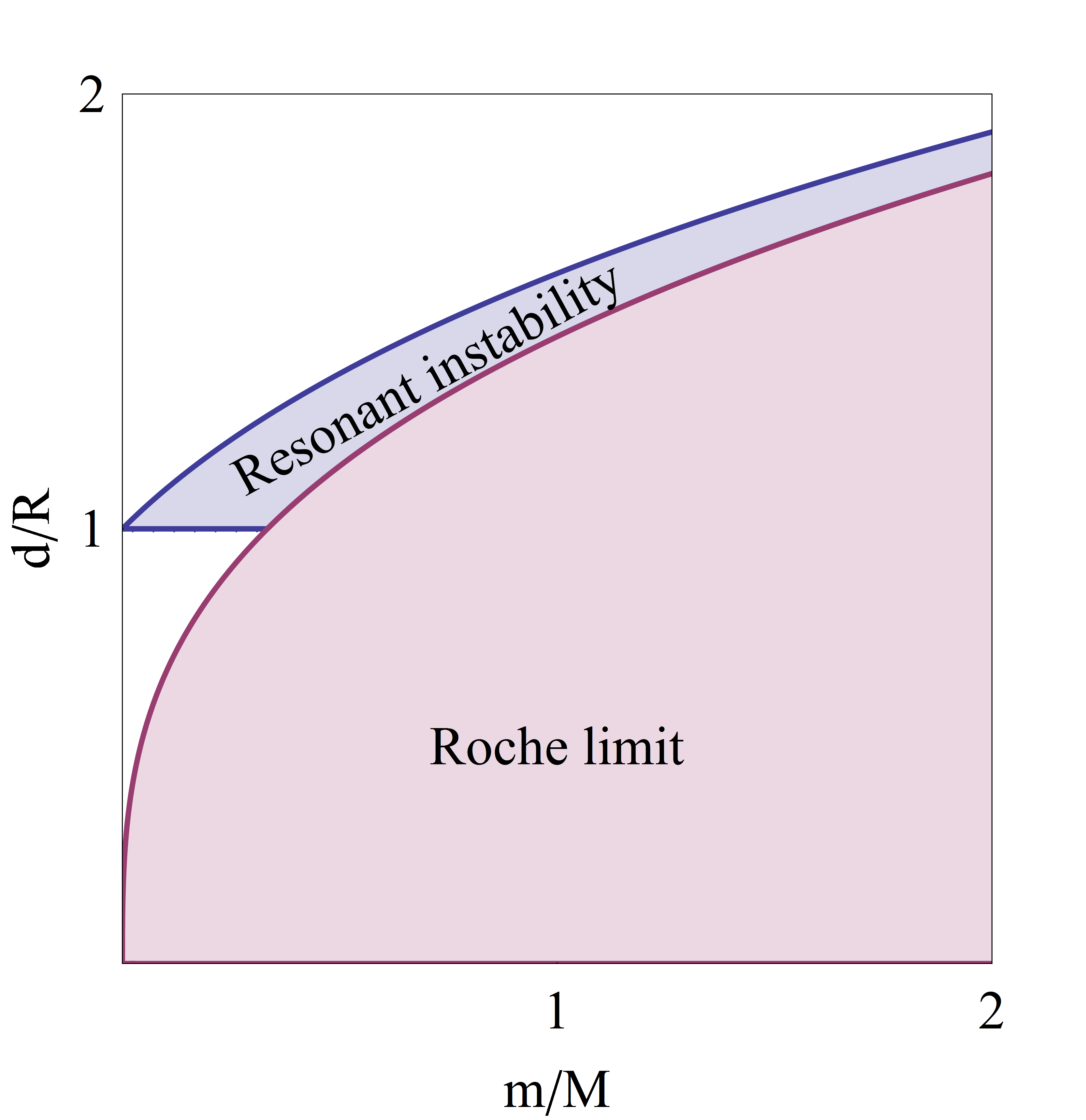}
\caption{\label{strip} Sketch of the resonant instability region given by
eq.~\eqref{roche_like}, with Roche limit, eq.~\eqref{roche}, superimposed on top. 
Blue strip marked "Resonant instability" shows parameter range, where phenomenon
already operates, but mass $M$ is still away from Roche boundary.
} 
\end{figure}

Roche stability limit can be put in the form:
\begin{equation}
\label{roche}
\frac{d}{R} > \kappa \left( \frac{m}{M} \right)^{1/3},
\end{equation}
where $\kappa=\sqrt[3]{3}$. Note, that for $m \gg M$
criteria \eqref{roche_like} and \eqref{roche} are
asymptotically identical. Resonant instability
\eqref{roche_like} operates before Roche limit,
leaving thin strip on $d/R - m/M$ plane, see Fig.~\ref{strip}.
Depending on actual value of $\kappa$ in \eqref{roche},
instability strip might widen if $\kappa < \sqrt[3]{3}$,
or disappear completely for large $m/M$ if $\kappa>\sqrt[3]{3}$.
Only for $m/M < 1$ instability is always present before
Roche limit. In other situations, mass $M$ might be destroyed
by tidal forces, before resonant instability
takes its time. The latter require at least
few orbital periods or more to be effective.

\subsection{Numerical verification of the instability and the long-term behavior in non-linear regime}
\label{Sect_numver}

\begin{figure*}
\includegraphics[width=0.3\textwidth]{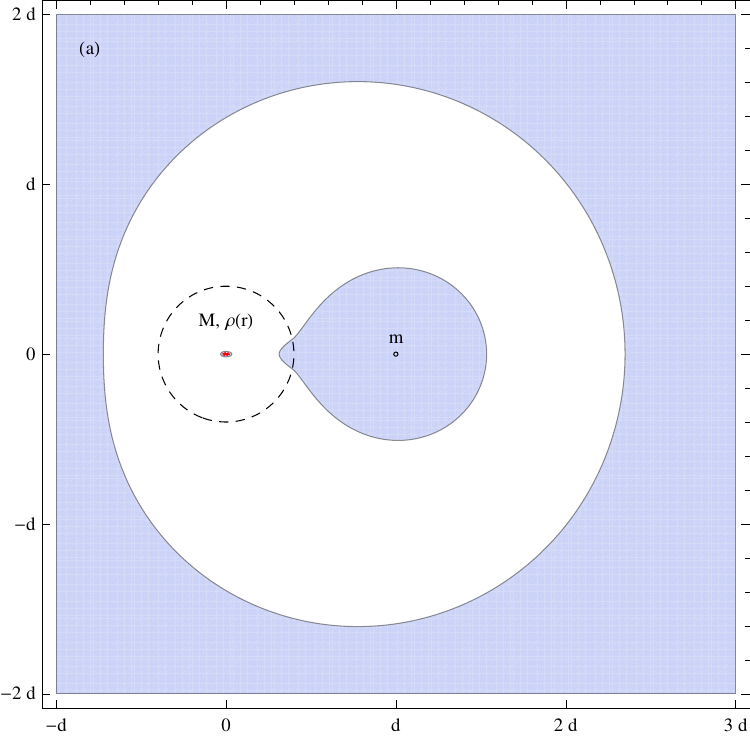}
\includegraphics[width=0.3\textwidth]{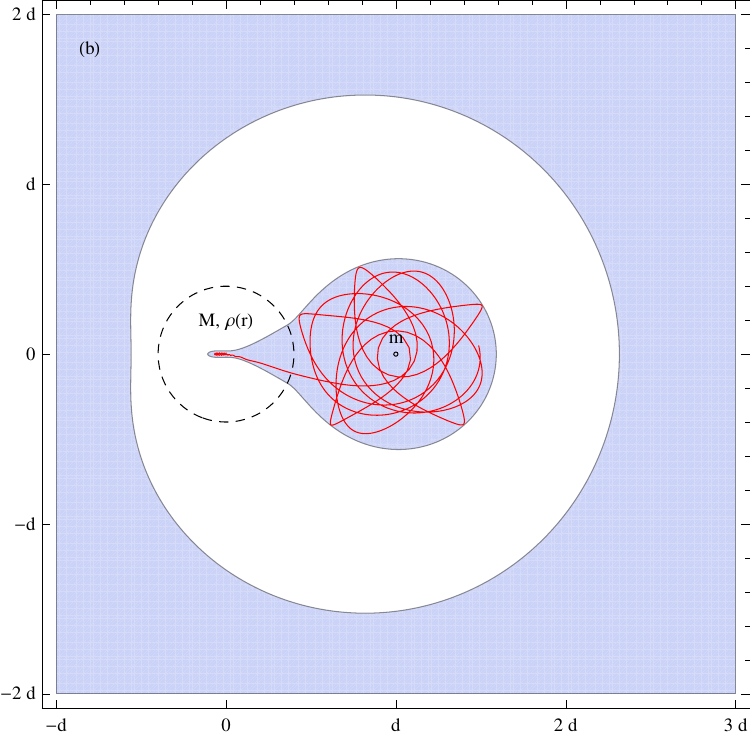}
\includegraphics[width=0.3\textwidth]{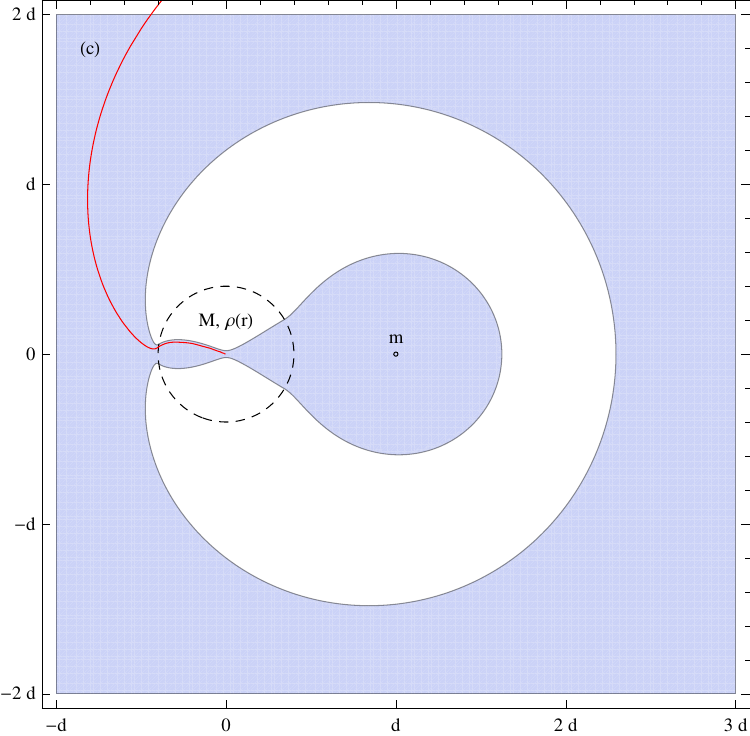}
\caption{\label{hillregion} Three typical cases of the dynamics: 
stable (a), unstable chaotic (b), unstable with ejection (c). Gravitational potential 
of the uniform density ball \eqref{phi_ball} has been used to plot shaded region
$U(x,y)<\delta E$ from \eqref{energy_ball}.
}
\end{figure*}

System \eqref{rpc3bsys} is a good tool to derive the instability 
criteria \eqref{crit}. For unstable cases point mass is likely to abandon central region. Whole density 
distribution $\rho(r)$, not just $\rho(0)$, becomes important. Location of ''surface radius'' defined as $\rho(R)=0$
determine cases of ''core ejection''. 
Non-constant density cause harmonic oscillator force change into more realistic, but also more complicated one. 
Introducing mass coordinate $m(r)$:
$$
m(r) = 4 \pi \int_0^r \rho(\zeta) \zeta^2 d \zeta
$$
the system of equations of motion becomes similar to \eqref{rpc3bsys}, but now
$k=k(x,y)$:
\begin{equation}
\label{variable_rho}
k = \frac{G \,m(r)}{r^3}- \omega^2, \quad r^2=x^2+y^2.
\end{equation}
System \eqref{rpc3bsys} with non-constant $k(x,y)$ given
by \eqref{variable_rho} is much more general, and
cover astrophysically interesting cases with non-constant
density.

Mass $M$ is now equal to:
$$
M \equiv m(R) =  4 \pi \int_0^R \rho(\zeta) \zeta^2 d \zeta.
$$
If we could allow mass $m$ to orbit inside region where $\rho(r)>0$, the inertial mass $M_{inert}$ 
would become different than the gravitational mass $M_{grav}$. In such a situation, 
circular two-body problem has slightly 
different solution compared to classic one. Kepler frequency in particular is given by:
\begin{equation}
\label{inertial_mass}
\omega^2 = G \frac{M_{grav}+m \frac{M_{grav}}{M_{inert}}}{d^3}.
\end{equation}
I do not consider consequences of \eqref{inertial_mass} further.
I assume, that $R<d$, i.e., mass $m$ orbits outside mass $M$.

For general $\rho(r)$ we still are able to derive conserved energy:
\begin{subequations}
\label{energy_ball}
\begin{equation}
E = \frac{1}{2} \dot{x}^2 + \frac{1}{2} \dot{y}^2 + U(x,y)
\end{equation}
\begin{equation}
\label{top_hill}
U = \phi(r) -\frac{1}{2} \omega^2 \, r^2 
 - \frac{ G\, m }{\sqrt{(x-d)^2+y^2}}+\frac{G\, m (x+d)}{d^2}
\end{equation}
\end{subequations}
where $r=\sqrt{x^2+y^2}$ and:
\begin{equation}
\label{aux_int}
\phi(r) = G \int_0^r \frac{m(\zeta)}{\zeta^2}  \; d \zeta
\end{equation}
is the gravitational potential of the star. For compatibility with \eqref{rpc3bsys} we 
choose $\phi(0)=0$.

Simplest example, uniform density ball of radius $R$,
leads to a piecewise-constant $\rho(r)$:
\begin{equation}
\label{rho_ball}
\rho(r) = 
\begin{cases}
\rho_0 \quad \text{for} \quad r<R\\
0 \quad \text{for} \quad r>R.
\end{cases}
\end{equation} 
Gravitational potential for \eqref{rho_ball} calculated from \eqref{aux_int} is:
\begin{equation}
\label{phi_ball}
\phi(r) = 
\begin{cases}
\frac{2}{3} \pi G \rho_0 r^2 \quad \text{for} \quad r<R\\
 - \frac{G M}{r} + \frac{3}{2} \frac{G M}{R}
 \quad \text{for} \quad r>R.
\end{cases}
\end{equation} 

Before presentation of numerical calculations, it is desirable to understand possible trajectories
qualitatively. Using energy considerations, we can find allowed regions
on $x-y$ plane. Test body at rest in the central point $x=0,y=0$ has an energy of $E=0$.
Three typical cases are presented for uniform sphere of density $\rho$ and radius $R$ 
(dashed circle in Fig.~\ref{hillregion}). 
Energy had been perturbed with small positive value $\delta E$. 
Allowed Hill region:
$$
U(x,y)<\delta E
$$ 
has been shaded. In stable situation, see Fig.~\ref{hillregion}, panel (a), we have three disconnected regions: 
central area of mass $M$, neighborhood of mass $m$, and ''outer space'' extending to the infinity. 
Perturbed test body simply oscillates with frequency related to the central density, simultaneously 
rotating like Foucault pendulum. In the presence of the drag it will settle down at the geometrical 
center of mass $M$. When the density drops to a value below critical, given by eq.~\eqref{crit_astro}, 
central region of mass $M$ and area surrounding mass $m$ become connected, cf. Fig.~\ref{hillregion}, 
middle panel (b). Test body begin to oscillate with growing amplitude forced by gravitational pull of 
mass $m$.  After some time, depending on amplitude of initial perturbation, test body is ejected from 
mass $M$ and enter chaotic orbit around mass $m$, see Fig.~\ref{hillregion}, panel (b). It is still 
bounded with binary system. With density much smaller than critical, all three allowed regions become 
connected, and the test body will be ejected from the system, spiraling out into infinity 
(Fig.~\ref{hillregion}, panel c). Described behavior has been confirmed by the numerical
solution of system \eqref{rpc3bsys} with non-constant $k$ from eq.~\eqref{variable_rho}. Example 
trajectories for uniform density ball are shown in Fig.~\ref{hillregion} using red curves.

Analyzed instability is not strictly resonant behavior with Kepler frequency \eqref{kepler_frequency}, 
because instability criterion \eqref{crit1} includes additional factor 3 before mass $m$. However, 
using language of resonances we can easily understand underlying physics \citep{3bodyresonance}. Consider 
classical first-year student textbook example of uniform density ball with radius $R$ 
and satellite in circular orbit just 
above surface ($d=R$). Well known result is that internal harmonic oscillator and orbital frequencies are 
identical. Therefore, test mass in the center is in resonance with small mass in orbit. However, 
astrophysical bodies are centrally condensed ($\rho_c>\bar{\rho}$) and binary companions are usually 
more distant ($d>R$), so this resonance disappear. But orbital frequency depends on both masses $M$ and $m$. 
By adding mass to the system, in form of larger mass $m$, we can make these frequencies equal again. 
We have never assumed that mass $M$ is dominant, so added mass $m$ might be arbitrarily large.

Inevitable presence of dynamic drag inside gaseous objects (including GC, \citealp[see][]{1997A&ARv...8....1M}) 
and other forms of friction 
forces acting on a test body inside mass $M$ has surprisingly only 
a minor effect on resulting dynamics, at least in the framework
of presented model. Drag suppress amplitude of oscillations and delay onset 
of the instability. However, in the resonant regime,   point $x=0,y=0$ 
has a ''top-hill'' position in the potential landscape given by the equation \eqref{top_hill}. 
In effect, the global dynamics remains the same. Numerical tests with simple drag of the form:
\begin{equation}
- \kappa(\rho)\, \dot{\mathbf{r}}
\end{equation}
confirmed this, and revealed, that main effect is to delay onset of the instability until ejection from the system becomes possible, 
cf.~Fig.~\ref{hillregion}, panel (c). Therefore, ejection is more likely outcome, 
while chaotic orbits (Fig.~\ref{hillregion}, panel (b)) might be rare. After test body leave mass $M$ evolution 
is identical to the classical restricted planar circular three-body problem. 

\subsection{Three body instability \label{instability_3d}}

In this section I have removed assumptions of planar motion and negligible mass of the third body.
Full system of nine equations describing motion of the three masses without additional simplifying assumptions 
can be transformed to the co-rotating system as well.  
Very lengthy calculations for full three body problem in three dimensions,
show that criteria \eqref{crit} nearly survive. Using quantifier elimination for stability 
analysis of the linearized 18-order eigensystem we are able to show, that
instability is present if:
\begin{subequations}
\label{critB}
\begin{equation}
\label{critB1}
\frac{M}{d^3} < \frac{4}{3} \pi \rho < \frac{ M + 3 m \left( 1+\mu/M \right)^{-1}}{d^3}
\end{equation}
\begin{equation}
\frac{4}{3} \pi \rho < \frac{1}{2} \frac{m}{d^3} \frac{M+\mu-m/8}{M+\mu+m} \left( 1+\mu/M \right)^{-1}.
\end{equation}
\end{subequations} 
New factor in \eqref{critB} is the mass of the third body equal to $\mu$.
For $\mu \to 0$, \eqref{critB} reduces to \eqref{crit}.

Again, only \eqref{critB1} is of astrophysical interest. Using orbital frequency:
$$
\omega^2 = \frac{G (m + \mu +M)}{d^3}
$$
and internal frequency:
$$
\omega_c^2 = \frac{4}{3} \, \pi G\; \rho
$$
we may write \eqref{critB1} as:
$$
\omega^2 + \frac{G (m + \mu)}{d^3}< \omega_c^2  < \omega^2 + \frac{G (m + \mu)}{d^3}+ \frac{3 G m}{d^3} \left( 1+\mu/M \right)^{-1}.
$$
Instability is a consequence of internal and orbital frequency overlap. Width of the resonance
is proportional to ,,forcing'' mass $m$, and reduced by factor dependent on mass ratio $\mu/M$. 
The most important is magnitude of mass $m$, because it increase orbital frequency allowing for resonance.
Simultaneously, it increase width of the instability window.

\section{Discussion of potential astrophysical sites of the instability}
\label{sites}

\subsection{Massive binaries}
\label{Subsect:redsupergiants}

Application of the results from Sect.~\ref{rpc3bp_sect} to a massive star is not straightforward,
because ,,core'' is not well-defined and separated from envelope. Red supergiants are indeed 
objects with nearly point-like core and extended low-density envelope. However, splitting radius is 
more or less arbitrary. It is also not clear what really will happen if instability becomes operational. 
Thorne-Zytkov objects an are exception, because central object is indeed well approximated by the point mass.

To overcome mentioned difficulties, I adopted the following procedure: star with total mass $M_{\ast}$ and 
radius $R_\ast$ is artificially divided into two parts:
(i) central ,,core'' region with $r<\xi$, and, (ii) outer envelope with $R_{\ast}>r>\xi$. Now, instability 
criteria \eqref{critB} are functions of parameter $\xi$ with:
\begin{equation}
\mu = \tilde{m}(\xi), \quad M = M_{\ast}-\tilde{m}(\xi), \quad \rho = \rho(\xi),
\end{equation}
where $\tilde{m}(\xi)$ denotes mass enclosed by sphere with radius $\xi$.
To further reduce complexity (dimensionality) of the analysis I assume that perturbing mass $m$ is as close 
to star as possible, i.e, $d=R_{\ast}$. Stellar model s15 of \cite{2002RvMP...74.1015W} with mass $M_{\ast}=12.8 M_\odot$  
and radius $R_\ast=3.85 AU$ at Si burning stage has been used as an example. The system is stable if:
\begin{equation}
\frac{4}{3} \pi \rho(\xi) R_{\ast}^3 >  \left( M_\ast + 3\, m \right)  \left( 1-\frac{\tilde{m}(\xi)}{M_\ast} \right).
\end{equation}
In the above example ,,unstable core'' has a minimal mass of $\simeq4.3$~$M_\odot$. Noteworthy, edge of 
the He core is placed at $\simeq 4.2~M_\odot$. This result do not significantly vary within mass range 
$1\;M_\odot<m<100\;M_\odot$.
I conclude that splitting the red giant into helium core and hydrogen envelope is the most appropriate. 
If we treat He core as a point mass, model presented in the article can be applied. It is likely, that 
during supernova event in close binary system, explosion engine will be displaced with respect to geometrical 
center of the hydrogen envelope. Recently discovered stripped helium core (low-mass white dwarf) 
in binary system \citep{2011MNRAS.418.1156M} might have been formed similar way.

\subsection{Exoplanets and moons}
\label{Subsect:exomoons}

Exoplanet formation and structure often consider compact core accreting
mass in the form of extended low-density envelope \citep{1538-4357-612-1-L73}. In this situation, dominant
is the mass $m$, i.e., the central star, and instability occur if:
\begin{equation}
\label{exoplanet}
\frac{\rho}{1\, g/cc} \left( \frac{T}{1\, day} \right)^2 < 0.057 \, \frac{M_{env}}{M_{tot}},
\end{equation}
where: $\rho$ is the density of the envelope, $M_{env}$ is the mass of accreted envelope and $M_{tot}$
is the total (core+envelope) mass of the protoplanet.

It is not surprising that all of the analyzed exoplanets are stable according to criterion \eqref{exoplanet}.
This is also true for so-called Ultra Short Period Planets \citep{2006Natur.443..534S} with orbital period 
less than a day. However, stability margin is often small. We may speculate that some 
of the ,,puffy planets'' \citep{2011ApJ...742...59H}, i.e., 
very low density Jupiter-like objects close to the central star, were formed in process 
involving ejection of the dense planetary core due to instability presented in Sect.~\ref{instability_3d}. 
Even if the instability do not lead to the core ejection due to, e.g., friction, dissipated energy might 
inflate the planet. This alone, however, does not explain lack of the rocky core.

\subsection{Globular cluster}
\label{Subsect:GC}

Galactic globular clusters are suspected to harbor intermediate-mass black hole in  
the center \citep{2013ApJ...776..118S, 2013degn.book.....M, 2014MNRAS.438..487D}.
Therefore we can use above model to check if central position is stable with respect to
perturbations caused by the Galaxy. Using database of \citet[2010 edition]{1996AJ....112.1487H} 
and eq.~\eqref{crit_astro} I have found
that only few of GC are unstable, namely Lynga 7, FSR 1735, Terzan 4, 2MS-GC01, 2MS-GC02,
BH 261, GLIMPSE02 and GLIMPSE01.  In simulations of \cite{2013A&A...558A.117L} 
instability apparently has not appeared. It is not surprising due to resonant character
of phenomena. It is unlikely to encounter such an instability for randomly chosen set of initial conditions.
Very interesting is case of Terzan 3. It is stable
if we put $m = 2 \times 10^{11}$ M$_\odot$, i.e. mass of the Galaxy 
without dark matter \citep{2012A&A...546A.126S}.
However, if we include dark matter it becomes unstable. Therefore, the search for central 
black hole in Terzan 3 might provide falsification test for amount of the dark matter in the Galaxy.
If the dark matter dominates mass of the Galaxy, IMBH in Terzan 3 must not exist.
This requires further investigation, because orbits of GC are usually not circular, and
Galaxy cannot be treated as a point mass. Another complication is caused by infinite 
radius of popular GC models, like Plummer sphere, see discussion related to \eqref{inertial_mass}.
More detailed investigation of GC with IMBH and N-body validation of the model from Subsect.~\ref{instability_3d}
is in progress.

\subsection{LHC black hole}
\label{Subsect:LHC}

It has been speculated that LHC or other future high-energy experiment might produce artificial black hole,
that do not explode immediately \textit{via} Hawking radiation 
\citep{2010JHEP...02..079C,2010JPhCS.237a2008B,2010PhRvD..81e7702G}. Such a black hole would settle 
at central region of the Earth and slowly consume our home planet. I have applied instability criterion 
\eqref{crit_astro} to the Earth-Moon and Earth-Sun
systems. Unfortunately, central position is stable by a wide margin in the sense of instability \eqref{crit}.

\section{Conclusions and discussion}
\label{final}

Generalized three body model has been analyzed using analytical and numerical techniques. Instability
of the point mass in the core-halo system was found. Analytical criteria \eqref{crit} 
and \eqref{critB} were derived using linearized system, 
and verified numerically.

Results were applied to  astrophysical binaries,
where one of the companions has a core-halo structure. Few possible sites for the instability
were discussed: massive red supergiants in binary system (Subsect.~\ref{Subsect:redsupergiants}),
formation of the exoplanets (Subsect.~\ref{Subsect:exomoons})
and globular clusters with intermediate black hole (Subsect.~\ref{Subsect:GC}). 
Model can be applied to a more exotic situations, e.g., Thorne-Zytkov objects or central 
black holes of astrophysical and artificial origin as well. 

%Importance of instability, 

Binary and multiple systems in astrophysics are rather a rule than exception, as well as core-halo 
structure of components, including dark matter halos and central black holes.
Therefore, instability presented in the article might be of a very common occurrence in the nature, influencing
formation and evolution of the astrophysical bodies and structures on various scales.

Derived results are very important. Simple analytical model provides input parameters 
for more advanced ones, numerical simulations in particular. Without such a guide, 
finding resonant behavior randomly
sweeping parameter range is improbable. This is especially important
for 3D simulations, which are limited by available computing power to just a few models
\citep{2013arXiv1312.3658H}.

%limitations of mechanical model

In the real world, instability appears in situations, where perturbing mass $m$ is either 
very close do mass $M$, or is much larger than mass $M$.
If the former case it is likely, that mass $m$ eventually enter into mass $M$, with dynamic drag causing inspiral,
sweeping all orbital frequencies. Encountering instability conditions seems inevitable. However, reduced
timescale available to system might prevent instability despite its exponential growth. 
If $m \gg M$ tidal effects are non-negligible, and instability asymptotically reduces to the Roche limit. 
Body with mass $M$ might 
be destroyed by tidal forces, before resonant instability becomes operational. However, resonant instability
appear before tidal disruption. Assumption of spherical symmetry and
neglected rotation might lead to some additional effects, but usually spherical models are good enough
to derive instability criteria.

%chances for observational verification

Observational verification of the instability would be difficult in stars and
planets, due to proximity of the Roche limit and complex hydrodynamics with
similar timescales. For globular clusters, model looks essentially
correct. Minor details like dynamic drag, non-circular orbits and finite size 
of the Galaxy are manageable, at least numerically. Unfortunately, 
existence of IMBH in the GC center is still under debate. Interesting option
is an experimental verification of the instability at space station, using manufactured
bodied in the form of, e.g., gas or liquid filled spheres or balloons.

%directions of future research

Three body model needs to be validated. In particular, astrophysical bodies of interest are not rigid 
(typically gaseous, liquid or composed of particles), and might not react to a driving force as a whole.
Ultimately, three dimensional hydrodynamic model with appropriate treatment of external gravity source, 
either analytical or numerical, should be used to verify instability in binary stars.
For globular clusters, N-body simulations provide a good framework to test model (work is in progress).

\bibliographystyle{elsarticle-harv}
\bibliography{3body_v2}

\begin{thebibliography}{41}
\expandafter\ifx\csname natexlab\endcsname\relax\def\natexlab#1{#1}\fi
\expandafter\ifx\csname url\endcsname\relax
  \def\url#1{\texttt{#1}}\fi
\expandafter\ifx\csname urlprefix\endcsname\relax\def\urlprefix{URL }\fi

\bibitem[{{Arnett} and {Meakin}(2011)}]{2011ApJ...733...78A}
{Arnett}, W.~D., {Meakin}, C., Jun. 2011. {Toward Realistic Progenitors of
  Core-collapse Supernovae}. \apj 733, 78--+.

\bibitem[{{Bleicher} and {Nicolini}(2010)}]{2010JPhCS.237a2008B}
{Bleicher}, M., {Nicolini}, P., Jun. 2010. {Large extra dimensions and small
  black holes at the LHC}. Journal of Physics Conference Series 237~(1),
  012008--+.

\bibitem[{{Cannon}(1993)}]{1993MNRAS.263..817C}
{Cannon}, R.~C., Aug. 1993. {Massive Thorne-{\.Z}ytkow Objects - Structure and
  Nucleosynthesis}. \mnras 263, 817--+.

\bibitem[{{Cannon} et~al.(1992){Cannon}, {Eggleton}, {Zytkow}, and
  {Podsiadlowski}}]{1992ApJ...386..206C}
{Cannon}, R.~C., {Eggleton}, P.~P., {Zytkow}, A.~N., {Podsiadlowski}, P., Feb.
  1992. {The structure and evolution of Thorne-Zytkow objects}. \apj 386,
  206--214.

\bibitem[{Capiñski and Zgliczyñski(2011)}]{0951-7715-24-5-002}
Capiñski, M.~J., Zgliczyñski, P., 2011. Transition tori in the planar
  restricted elliptic three-body problem. Nonlinearity 24~(5), 1395.
\newline\urlprefix\url{http://stacks.iop.org/0951-7715/24/i=5/a=002}

\bibitem[{{Casadio} et~al.(2010){Casadio}, {Fabi}, {Harms}, and
  {Micu}}]{2010JHEP...02..079C}
{Casadio}, R., {Fabi}, S., {Harms}, B., {Micu}, O., Feb. 2010. {Theoretical
  survey of tidal-charged black holes at the LHC}. Journal of High Energy
  Physics 2, 79--+.

\bibitem[{{Chandrasekhar}(1963)}]{1963ApJ...138.1182C}
{Chandrasekhar}, S., Nov. 1963. {The Equilibrium and the Stability of the Roche
  Ellipsoids.} \apj 138, 1182.

\bibitem[{{Clayton} et~al.(1975){Clayton}, {Newman}, and
  {Talbot}}]{1975ApJ...201..489C}
{Clayton}, D.~D., {Newman}, M.~J., {Talbot}, Jr., R.~J., Oct. 1975. {Solar
  models of low neutrino-counting rate - The central black hole}. \apj 201,
  489--493.

\bibitem[{{den Brok} et~al.(2014){den Brok}, {van de Ven}, {van den Bosch}, and
  {Watkins}}]{2014MNRAS.438..487D}
{den Brok}, M., {van de Ven}, G., {van den Bosch}, R., {Watkins}, L., Feb.
  2014. {The central mass and mass-to-light profile of the Galactic globular
  cluster M15}. \mnras 438, 487--493.

\bibitem[{{Feldmeier} et~al.(2013){Feldmeier}, {L{\"u}tzgendorf}, {Neumayer},
  {Kissler-Patig}, {Gebhardt}, {Baumgardt}, {Noyola}, {de Zeeuw}, and
  {Jalali}}]{2013A&A...554A..63F}
{Feldmeier}, A., {L{\"u}tzgendorf}, N., {Neumayer}, N., {Kissler-Patig}, M.,
  {Gebhardt}, K., {Baumgardt}, H., {Noyola}, E., {de Zeeuw}, P.~T., {Jalali},
  B., Jun. 2013. {Indication for an intermediate-mass black hole in the
  globular cluster NGC 5286 from kinematics}. \aap 554, A63.

\bibitem[{{Gingrich}(2010)}]{2010PhRvD..81e7702G}
{Gingrich}, D.~M., Mar. 2010. {Production of tidal-charged black holes at the
  Large Hadron Collider}. \prd 81~(5), 057702--+.

\bibitem[{{Gu} et~al.(2003){Gu}, {Lin}, and
  {Bodenheimer}}]{2003ApJ...588..509G}
{Gu}, P.-G., {Lin}, D.~N.~C., {Bodenheimer}, P.~H., May 2003. {The Effect of
  Tidal Inflation Instability on the Mass and Dynamical Evolution of Extrasolar
  Planets with Ultrashort Periods}. \apj 588, 509--534.

\bibitem[{{Haghighipour}(2011)}]{2011ConPh..52..403H}
{Haghighipour}, N., Sep. 2011. {Super-Earths: a new class of planetary bodies}.
  Contemporary Physics 52, 403--438.

\bibitem[{{Handy} et~al.(2013){Handy}, {Plewa}, and
  {Odrzywolek}}]{2013arXiv1312.3658H}
{Handy}, T., {Plewa}, T., {Odrzywolek}, A., Dec. 2013. {Toward connecting
  core-collapse supernova theory with observations: I. Shock revival in a 15
  Msun blue supergiant progenitor with SN 1987A energetics}. ArXiv e-prints.

\bibitem[{{Harris}(1996)}]{1996AJ....112.1487H}
{Harris}, W.~E., Oct. 1996. {A Catalog of Parameters for Globular Clusters in
  the Milky Way}. \aj 112, 1487.

\bibitem[{{Hartman} et~al.(2011){Hartman}, {Bakos}, {Torres}, {Latham},
  {Kov{\'a}cs}, {B{\'e}ky}, {Quinn}, {Mazeh}, {Shporer}, {Marcy}, {Howard},
  {Fischer}, {Johnson}, {Esquerdo}, {Noyes}, {Sasselov}, {Stefanik},
  {Fernandez}, {Szklen{\'a}r}, {L{\'a}z{\'a}r}, {Papp}, and
  {S{\'a}ri}}]{2011ApJ...742...59H}
{Hartman}, J.~D., {Bakos}, G.~{\'A}., {Torres}, G., {Latham}, D.~W.,
  {Kov{\'a}cs}, G., {B{\'e}ky}, B., {Quinn}, S.~N., {Mazeh}, T., {Shporer}, A.,
  {Marcy}, G.~W., {Howard}, A.~W., {Fischer}, D.~A., {Johnson}, J.~A.,
  {Esquerdo}, G.~A., {Noyes}, R.~W., {Sasselov}, D.~D., {Stefanik}, R.~P.,
  {Fernandez}, J.~M., {Szklen{\'a}r}, T., {L{\'a}z{\'a}r}, J., {Papp}, I.,
  {S{\'a}ri}, P., Nov. 2011. {HAT-P-32b and HAT-P-33b: Two Highly Inflated Hot
  Jupiters Transiting High-jitter Stars}. \apj 742, 59.

\bibitem[{{Hawking}(1971)}]{1971MNRAS.152...75H}
{Hawking}, S., 1971. {Gravitationally collapsed objects of very low mass}.
  \mnras 152, 75--+.

\bibitem[{{Hur{\'e}} and {Dieckmann}(2012)}]{2012A&A...541A.130H}
{Hur{\'e}}, J.-M., {Dieckmann}, A., May 2012. {A substitute for the singular
  Green kernel in the Newtonian potential of celestial bodies}. \aap 541, A130.

\bibitem[{{Kuchner}(2003)}]{2003ApJ...596L.105K}
{Kuchner}, M.~J., Oct. 2003. {Volatile-rich Earth-Mass Planets in the Habitable
  Zone}. \apjl 596, L105--L108.

\bibitem[{{Landau} and {Lifshitz}(1969)}]{1969mech.book.....L}
{Landau}, L.~D., {Lifshitz}, E.~M., 1969. {Mechanics}. Oxford: Pergamon Press,
  1969, 2nd ed.

\bibitem[{Laughlin et~al.(2004)Laughlin, Bodenheimer, and
  Adams}]{1538-4357-612-1-L73}
Laughlin, G., Bodenheimer, P., Adams, F.~C., 2004. The core accretion model
  predicts few jovian-mass planets orbiting red dwarfs. The Astrophysical
  Journal Letters 612~(1), L73.
\newline\urlprefix\url{http://stacks.iop.org/1538-4357/612/i=1/a=L73}

\bibitem[{{L{\'e}ger} et~al.(2004){L{\'e}ger}, {Selsis}, {Sotin}, {Guillot},
  {Despois}, {Mawet}, {Ollivier}, {Lab{\`e}que}, {Valette}, {Brachet},
  {Chazelas}, and {Lammer}}]{2004Icar..169..499L}
{L{\'e}ger}, A., {Selsis}, F., {Sotin}, C., {Guillot}, T., {Despois}, D.,
  {Mawet}, D., {Ollivier}, M., {Lab{\`e}que}, A., {Valette}, C., {Brachet}, F.,
  {Chazelas}, B., {Lammer}, H., Jun. 2004. {A new family of planets?
  ``Ocean-Planets''}. Icarus 169, 499--504.

\bibitem[{Liska and Steinberg(1993)}]{Liska01011993}
Liska, R., Steinberg, S., 1993. Applying quantifier elimination to stability
  analysis of difference schemes. The Computer Journal 36~(5), 497--503.
\newline\urlprefix\url{http://comjnl.oxfordjournals.org/content/36/5/497.abstr%
act}

\bibitem[{{L{\"u}tzgendorf} et~al.(2013{\natexlab{a}}){L{\"u}tzgendorf},
  {Baumgardt}, and {Kruijssen}}]{2013A&A...558A.117L}
{L{\"u}tzgendorf}, N., {Baumgardt}, H., {Kruijssen}, J.~M.~D., Oct.
  2013{\natexlab{a}}. {N-body simulations of globular clusters in tidal fields:
  Effects of intermediate-mass black holes}. \aap 558, A117.

\bibitem[{{L{\"u}tzgendorf} et~al.(2013{\natexlab{b}}){L{\"u}tzgendorf},
  {Kissler-Patig}, {Gebhardt}, {Baumgardt}, {Noyola}, {de Zeeuw}, {Neumayer},
  {Jalali}, and {Feldmeier}}]{2013MmSAI..84..129L}
{L{\"u}tzgendorf}, N., {Kissler-Patig}, M., {Gebhardt}, K., {Baumgardt}, H.,
  {Noyola}, E., {de Zeeuw}, P.~T., {Neumayer}, N., {Jalali}, B., {Feldmeier},
  A., 2013{\natexlab{b}}. {Intermediate-mass black holes in globular clusters
  }. \memsai 84, 129.

\bibitem[{{L{\"u}tzgendorf} et~al.(2013{\natexlab{c}}){L{\"u}tzgendorf},
  {Kissler-Patig}, {Gebhardt}, {Baumgardt}, {Noyola}, {de Zeeuw}, {Neumayer},
  {Jalali}, and {Feldmeier}}]{2013A&A...552A..49L}
{L{\"u}tzgendorf}, N., {Kissler-Patig}, M., {Gebhardt}, K., {Baumgardt}, H.,
  {Noyola}, E., {de Zeeuw}, P.~T., {Neumayer}, N., {Jalali}, B., {Feldmeier},
  A., Apr. 2013{\natexlab{c}}. {Limits on intermediate-mass black holes in six
  Galactic globular clusters with integral-field spectroscopy}. \aap 552, A49.

\bibitem[{{L{\"u}tzgendorf} et~al.(2013{\natexlab{d}}){L{\"u}tzgendorf},
  {Kissler-Patig}, {Neumayer}, {Baumgardt}, {Noyola}, {de Zeeuw}, {Gebhardt},
  {Jalali}, and {Feldmeier}}]{2013A&A...555A..26L}
{L{\"u}tzgendorf}, N., {Kissler-Patig}, M., {Neumayer}, N., {Baumgardt}, H.,
  {Noyola}, E., {de Zeeuw}, P.~T., {Gebhardt}, K., {Jalali}, B., {Feldmeier},
  A., Jul. 2013{\natexlab{d}}. {M$_{•}$ - {$\sigma$}relation for
  intermediate-mass black holes in globular clusters}. \aap 555, A26.

\bibitem[{{Mardling}(2008{\natexlab{a}})}]{2008LNP...760...59M}
{Mardling}, R.~A., 2008{\natexlab{a}}. {Resonance, Chaos and Stability: The
  Three-Body Problem in Astrophysics}. In: {S.~J.~Aarseth, C.~A.~Tout, \&
  R.~A.~Mardling} (Ed.), The Cambridge N-Body Lectures. Vol. 760 of Lecture
  Notes in Physics, Berlin Springer Verlag. pp. 59--96.

\bibitem[{{Mardling}(2008{\natexlab{b}})}]{3bodyresonance}
{Mardling}, R.~A., 2008{\natexlab{b}}. Resonsnce, chaos and stability: The
  three-body problem in astrophysics. Lecture Notes in Physics 760, 59--96.

\bibitem[{{Maxted} et~al.(2011){Maxted}, {Anderson}, {Burleigh}, {Collier
  Cameron}, {Heber}, {G{\"a}nsicke}, {Geier}, {Kupfer}, {Marsh}, {Nelemans},
  {O'Toole}, {{\O}stensen}, {Smalley}, and {West}}]{2011MNRAS.418.1156M}
{Maxted}, P.~F.~L., {Anderson}, D.~R., {Burleigh}, M.~R., {Collier Cameron},
  A., {Heber}, U., {G{\"a}nsicke}, B.~T., {Geier}, S., {Kupfer}, T., {Marsh},
  T.~R., {Nelemans}, G., {O'Toole}, S.~J., {{\O}stensen}, R.~H., {Smalley}, B.,
  {West}, R.~G., Dec. 2011. {Discovery of a stripped red giant core in a bright
  eclipsing binary system}. \mnras 418, 1156--1164.

\bibitem[{{Merritt}(2013)}]{2013degn.book.....M}
{Merritt}, D., Jul. 2013. {Dynamics and Evolution of Galactic Nuclei}.
  Princeton University Press.

\bibitem[{{Meylan} and {Heggie}(1997)}]{1997A&ARv...8....1M}
{Meylan}, G., {Heggie}, D.~C., 1997. {Internal dynamics of globular clusters}.
  \aapr 8, 1--143.

\bibitem[{{Sahu} et~al.(2006){Sahu}, {Casertano}, {Bond}, {Valenti}, {Ed
  Smith}, {Minniti}, {Zoccali}, {Livio}, {Panagia}, {Piskunov}, {Brown},
  {Brown}, {Renzini}, {Rich}, {Clarkson}, and {Lubow}}]{2006Natur.443..534S}
{Sahu}, K.~C., {Casertano}, S., {Bond}, H.~E., {Valenti}, J., {Ed Smith}, T.,
  {Minniti}, D., {Zoccali}, M., {Livio}, M., {Panagia}, N., {Piskunov}, N.,
  {Brown}, T.~M., {Brown}, T., {Renzini}, A., {Rich}, R.~M., {Clarkson}, W.,
  {Lubow}, S., Oct. 2006. {Transiting extrasolar planetary candidates in the
  Galactic bulge}. \nat 443, 534--540.

\bibitem[{{Sikora} et~al.(2012){Sikora}, {Bratek}, {Ja{\l}ocha}, and
  {Kutschera}}]{2012A&A...546A.126S}
{Sikora}, S., {Bratek}, {\L}., {Ja{\l}ocha}, J., {Kutschera}, M., Oct. 2012.
  {Gravitational microlensing as a test of a finite-width disk model of the
  Galaxy}. \aap 546, A126.

\bibitem[{Strzebonski(2000)}]{strzebonski}
Strzebonski, A., 2000. Solving algebraic inequalities. Mathematica J. 7,
  525--541.

\bibitem[{{Sun} et~al.(2013){Sun}, {Jin}, {Gu}, {Liu}, {Lin}, and
  {Lu}}]{2013ApJ...776..118S}
{Sun}, M.-Y., {Jin}, Y.-L., {Gu}, W.-M., {Liu}, T., {Lin}, D.-B., {Lu}, J.-F.,
  Oct. 2013. {Do Intermediate-mass Black Holes Exist in Globular Clusters?}
  \apj 776, 118.

\bibitem[{{Thorne} and {Zytkow}(1977)}]{1977ApJ...212..832T}
{Thorne}, K.~S., {Zytkow}, A.~N., Mar. 1977. {Stars with degenerate neutron
  cores. I - Structure of equilibrium models}. \apj 212, 832--858.

\bibitem[{{Umbreit} and {Rasio}(2013)}]{2013ApJ...768...26U}
{Umbreit}, S., {Rasio}, F.~A., May 2013. {Constraining Intermediate-mass Black
  Holes in Globular Clusters}. \apj 768, 26.

\bibitem[{{Vanture} et~al.(1999){Vanture}, {Zucker}, and
  {Wallerstein}}]{1999ApJ...514..932V}
{Vanture}, A.~D., {Zucker}, D., {Wallerstein}, G., Apr. 1999. {Is U Aquarii a
  Thorne-{\.Z}ytkow Object?} \apj 514, 932--938.

\bibitem[{Wang and Xu(2013)}]{wang2013gravity}
Wang, Y., Xu, S., 2013. Gravity gradient torque of spacecraft orbiting
  asteroids. Aircraft Engineering and Aerospace Technology 85~(1), 8--8.

\bibitem[{{Woosley} et~al.(2002){Woosley}, {Heger}, and
  {Weaver}}]{2002RvMP...74.1015W}
{Woosley}, S.~E., {Heger}, A., {Weaver}, T.~A., Nov. 2002. {The evolution and
  explosion of massive stars}. Reviews of Modern Physics 74, 1015--1071.

\end{thebibliography}

\end{document}